# Evidence of electronic phase arrest and glassy ferromagnetic behaviour in $(Nd_{0.4}Gd_{0.3})Sr_{0.3}MnO_3$ manganite : Comparative study between bulk and nanometric samples

S. Kundu[a] and T. K. Nath[*]

*Department of Physics and Meteorology, Indian Institute of Technology, Kharagpur, West Bengal 721302, India*
[a]*email:* souravphy@gmail.com

## Abstract

*The effect of doping of rare earth Gd* $^{3+}$ *ion replacing Nd* $^{3+}$ *in* $Nd_{0.7}Sr_{0.3}MnO_3$ *is investigated in details. Measurements of resistivity, magnetoresistance, magnetization, linear and non linear ac magnetic susceptibility on chemically synthesized* $(Nd_{0.7-x}Gd_x)Sr_{0.3}MnO_3$ *shows various interesting features with doping level x=0.3. Comparative study has been carried out between a bulk and a nanometric sample (grain size ~ 60 nm) synthesized from the same as prepared powder to maintain identical stoichiometry. Resistivity of the samples shows strong dependence on the magnetic field – temperature history. The magnetoresistance of the samples also show strong irreversibility with respect to sweeping of the field between highest positive and negative values. Moreover, resistivity is found to increase with time after field cooling and then switching off the field. All these phenomena have been attributed to phase separation effect and arrest of phases in the samples. Furthermore, the bulk sample displays a spin glass like behaviour as evident from frequency dependence of linear ac magnetic susceptibility and critical divergence of the nonlinear ac magnetic susceptibility. The experimentally obtained characteristic time* $\tau_0$ *after dynamical scaling analysis of the frequency dependence of the ac susceptibility is found to be* $10^{-17}$ *s which implies that the system is different from a canonical spin glass. An unusual frequency dependence of the second harmonic of ac susceptibility around the magnetic transition temperature led us to designate the magnetic state of the sample to be glassy ferromagnetic. On reduction of grain size low field magnetoresistance and phase arrest phenomena are found to enhance but the glassy state is observed to be destabilized in the nanometric sample.*

---

[*]Corresponding author: tnath@phy.iitkgp.ernet.in

Tel: +91-3222-283862
Area code: 721302,
INDIA

# I. Introduction

The manganites with perovskite $ABO_3$ structure have attracted enormous research interest over the years because of their very rich and unusual physical properties which are easily tuneable through various ways like substitution/doping, external pressure, change in grain size etc [1]. Firstly, it has been found that the appropriate substitution of the A-site trivalent elements with divalent materials (like $Sr^{2+}$ and $Ca^{2+}$ etc.) introduces ferromagnetism and metallicity accompanied by a colossal magnetoresistance effect governed mainly by the so called double exchange mechanism [2]. Furthermore, it has also been found that substitution of the trivalent A- site with a different trivalent ion (like the rare earth elements), in a double exchange ferromagnetic manganite introduces many other fascinating effects on magnetic and electronic properties [3-5]. Such A-site substitution has some fundamental effects on the system as mentioned below. Such as, substitution by an ion with different ionic radius changes the average A-site ionic radius ($\langle r_A \rangle$) of the parent manganites effecting directly the band width of the material [3,6]. This type of substitution also promotes A-site disorder due to random distribution of the A-site cations with different size and generally estimated by the variance ($\sigma^2 = \sum y_i r_i^2 - \langle r_A \rangle^2$ , $y_i$ is the fractional occupancy of the $i^{th}$ ion in the A-site) [4,7]. It has been observed that above mentioned doping has profound influence on magnetic transition temperature as well as on other magnetic and electronic properties [3,8-10]. Moreover, if the dopant has magnetic moment then an additional magnetic coupling takes place between the Mn site and the dopant apart from the $Mn^{3+}$ /$Mn^{4+}$ double exchange interaction [11,12]. Ultimately, the phase separation effect which plays a significant role towards most of the behaviour of manganites, is enhanced on such substitution [10]. A large amount of study has been carried out on doping of rare earth elements in the A- site of La based manganites (La-Sr/Ca-Mn-O). These parent compounds are known to be excellent double exchange system with non magnetic A-site ($La^{3+}$). In many cases, it has been found that doping with rare earth ions influences and enhances the colossal magnetoresistive properties of the systems [3,8,13] besides having strong effect on other magnetic and electronic properties. However, most of the studies were performed on bulk manganite systems. At this stage, we perform a detailed study of doping of the rare earth $Gd^{3+}$ ion having a lower ionic radius and higher magnetic moment than $Nd^{3+}$ in the A-site of an intermediate bandwidth double exchange system $Nd_{0.7}Sr_{0.3}MnO_3$. So far as the bandwidth is concerned this system stands in between La-Sr-Mn-O and La-Ca-Mn-O and hence interesting candidate for study in regards to doping on A-site. We observe the interesting phenomena of thermo-magnetic history dependent electronic - transport and strong relaxation of resistivity with time with a doping level of 30 % of $Gd^{3+}$ $(Nd_{0.7-x}Gd_x)Sr_{0.3}MnO_3$ with x = 0.3). Moreover, we observe evidence of spin glass like behaviour in this system. None of these above mentioned properties of $Nd_{0.4}Gd_{0.3}Sr_{0.3}MnO_3$ ($\langle r_A \rangle$ = 1.198 Å) have been found in the parent compound $Nd_{0.7}Sr_{0.3}MnO_3$ ($\langle r_A \rangle$ = 1.206 Å) and shows the strong influence of doping effect in manganites. Apart from study in the bulk system, we have also synthesized the nanometric counterpart of the same sample and performed the similar measurements on it. Reduction of grain size is found to have strong influence in both electronic and magnetic properties of the system. The present study also includes a detailed comparative study of the observed properties in bulk and nanometric samples with same stoichiometry.

## II. Experimental Details

Polycrystalline samples of $Nd_{0.4}Gd_{0.3}Sr_{0.3}MnO_3$ have been synthesized through chemical pyrophoric reaction route [14]. High purity $Nd_2O_3$, $Gd_2O_3$, $Sr(NO_3)_2$ and $Mn(CH_3COO)_2$ were dissolved in water with appropriate amount of $HNO_3$. Triethanolamine was then added to this solution and stirred continuously while heating the mixture at $180^0$C.

Finally, combustion takes place producing a black fluppy powder which, on calcination, produces polycrystalline samples of $Nd_{0.4}Gd_{0.3}Sr_{0.3}MnO_3$. To synthesize nanometric as well as bulk samples of same stoichiometry we have calcinated one part of the as prepared sample at $750^0$C and the remaining at $1150^0$C. The former produces nanometric samples of $Nd_{0.4}Gd_{0.3}Sr_{0.3}MnO_3$ (NGSMO_N), while the latter produces the bulk sample of $Nd_{0.4}Gd_{0.3}Sr_{0.3}MnO_3$ (NGSMO_B).

The structural characterization of the samples has been carried out through High Resolution x-ray Diffraction (HRXRD) from Panalytical, High Resolution Transmission Electron Microscopy (HRTEM) from JEOL Ltd., Japan and Field Emission Scanning Electron Microscopy (FESEM) from Carl Zeiss, Germany. Electronic – transport properties of the samples was measured employing standard linear four probe technique on pellets (approximate dimension ~ 8 mm×3 mm×0.3 mm) of the samples down to a lowest temperature of 2 K and under a highest magnetic field of 8 T. We have employed conductive silver paint for making contacts on the sample surface. Measurement of resistivity and magnetoresistance has been carried out by applying low dc current (employing Keithley-6221 current source) of magnitude 10 μA for NGSMO_B and 1 μA for NGSMO_N sample to avoid Joule heating in the samples. A closed cycle Helium refrigeration cryostat attached with an 8 T superconducting magnet from Cryogenics Ltd., U. K. were employed to perform these measurements. The static (dc) magnetization, M (T) and M (H), were measured in a superconducting quantum interference device (SQUID) from Quantum Design**.** The linear and nonlinear ac susceptibility was measured down to 2 K employing a homemade ac susceptibility coil fitted inside the Helium refrigeration cryostat. A Lock -In -Amplifier (SR830) was used to measure the induced signal. Temperature of the sample was controlled by high precision PID temperature controllers from Lakeshore Cryotronics, USA (model 325 and 340) with temperature stability better than ±50 mK.

## III. Results

### 1. Structural properties

The HRXRD data shown in the Fig. 1 (a) reveals that the samples are of single phase with no detectable impurity phase present. For further structural analysis we have performed Rietveld refinement of the experimental data using Maud programme. Both the nanometric and the bulk samples are found to crystallize in orthorhombic *Pnma* space group. The refined lattice parameters are 5.4401(8) Å (a), 7.6841(4) Å (b) and 5.4630(5) Å (c) for NGSMO_B sample. The lattice parameters for NGSMO_N sample are 5.4584(3) Å (a), 7.6551(6) Å (b) and 5.4341(10) Å (c). The unit cell volume of NGSMO_B sample (228.3 $Å^3$) is found to reduce slightly in the NGSMO_N sample (227.1 $Å^3$). Hence, we see that the structural change is not significant with the change of grain size. The HRTEM (Fig. 1 (b)) image of NGSMO_N sample displays that the grain size of the sample is in the nanometric regime (less than 100 nm). The average grain size is about 60 nm as determined from the plot of distribution (Fig. 1(c)). On the other hand, the FESEM image of NGSMO_B (Fig.1 (d)) shows micro meter (µm) size grains and the sample is expected to show bulk properties. The spotty rings in the SAED pattern in Fig. 1 (e) of NGSMO_N sample indicate polycrystalline nature of this sample.

### 2. Transport properties

A detailed analysis of the electronic - and magneto - transport properties of both the samples has been carried out. The resistivity as a function of temperature of the samples has been measured at zero magnetic field and shown in the Fig. 2. Both the samples show a metal to semiconductor/insulator transition followed by a resistivity minima and resistivity upturn in the low temperature regime. The striking difference between these two samples is the presence of a huge upturn of resistivity for NGSMO_N (Fig. 2 (a)). In case of NGSMO_B sample this upturn is not so prominent (Fig. 2(b)). This indicates a strong influence of grain size on the observed low temperature minima of polycrystalline manganites. The temperature dependent resistivity in the low temperature regime has been analyzed in more details. Such observation of resistivity minimum is generally explained in terms of Coulomb blockade [15], Kondo effect [16,17] or electron-electron (e-e) interaction [16,17]. Due to Coulomb blockade effect resistivity goes as $\exp(\sqrt{(\Delta/T)})$ in the low temperature regime. Here $\Delta$ is equivalent to charging energy. The Kondo effect, which is generally observed in low resistive magnetic alloys exhibits ln(T) dependence of resistivity in the low temperature regime. In disordered, high resistive system the elastic type e-e scattering occurs when the mean free path of the electrons becomes small and resistivity shows $T^{1/2}$ dependence. With the increase of temperature inelastic scattering of electrons increases and gives rise to the resistivity minimum. This inelastic scattering is assumed here to have a $T^n$ dependence, in general. Coulomb blockade type model is found to produce very poor fitting of the resistivity data in the low temperature regime of NGSMO_B sample (at H = 0 T). Then we employ equations like $\rho = \rho_0 - \rho_s \ln T + \rho_{in} T^n$ assuming Kondo effect or $\rho = \rho_0 - \rho_e T^{1/2} + \rho_{in} T^n$ assuming e-e interaction to fit the low temperature resistivity data of NGSMO_B sample as shown in the inset of Fig. 2(b). One can easily observe that the resistivity minimum in this sample is best described through e-e interaction process as the latter equation gives a better fit. The obtained parameters $\rho_0$, $\rho_e$, $\rho_{in}$ and n are found to be 563 ohm-cm, 72.5 ohm-cm-$K^{-1/2}$, 7.3E-4 ohm-cm-$K^{-n}$ and 3.4, respectively. We employ similar approach to fit the low temperature resistivity data of NGSMO_N sample. However, none of the above mention models are found to fit the experimental data satisfactorily. This indicates that the mechanism behind the huge low temperature upturn and minima in resistivity of NGSMO_N sample is more complicated and different from that in NGSMO_B sample. We primarily attribute this complicacy in NGSMO_N sample to the presence of enhanced grain boundary where the actual nature of electron transport is not understood so far. Interestingly, NGSMO_N sample shows a thermal hysteresis around the metal insulator transition as shown in the inset of Fig. 2 (a). This thermal hysteresis is a generic feature of first order phase transition [18, 19].

During the measurement of resistivity of the samples under magnetic field/without magnetic field, we adopted mainly four different protocols.

1. The resistivity was measured during heating at zero magnetic field after cooling the sample without the application of field (ZFC_W resistivity).
2. The sample was cooled in zero magnetic field and the data was taken during heating at specific selected magnetic field (ZFC_FW).
3. The sample was cooled under selected magnetic field and the data was taken during heating at the same field (FC_FW).
4. The sample was cooled under selected magnetic field and data was taken during heating at zero applied field (FC_W).

In all these cases the cooling is always started from a temperature much above the metal – insulator transition temperature.

The methods described in 2 and 3 are similar to what is generally followed in field cooled and zero field cooled magnetization measurements, magnetization being replaced by resistivity of the sample. The method described in 4 is possible as magnetic field is not necessary for measurement of resistivity.

Fig. 3 (a) displays the ZFC_FW and FC_FW resistivity curves for the NGSMO_N sample. One can observe that the resistivity vs. temperature curves following these two methods at a particular field tends to bifurcate from each other at a very low temperature (<20 K). At the lowest applied field (0.1T) the bifurcation is not prominent. However, at 1T field the bifurcation seems to be the maximum and decreases with increasing field thereafter. As a consequence, the bifurcation point shifts to lower value of temperature in the high field regime. The FC_FW curve remains below the ZFC_FW curve at all applied fields. This precisely implies that the sample remembers the history even during heating at the same field which is reflected in the bifurcation. A history dependence of resistivity of a sample is not a very common phenomenon and not commonly reported in literature especially in case of manganites with optimally doped regime. Even for the parent compound $Nd_{0.7}Sr_{0.3}MnO_3$, no such history dependence of resistivity is reported to the best of our knowledge. So, the observed effects can be attributed to the doping of $Gd^{3+}$ in the A-site of the system under study. A similar measurement has been carried out on NGSMO_B sample to compare the effects with the nanometric counterpart. A similar bifurcation between the ZFC_FW and FC_FW curves has been found as shown in the Fig. 3 (b). Obviously, in case of NGSMO_B the bifurcation is small compared to that in the NGSMO_N sample. We, hence, can conclusively state that the thermo-magnetic history dependence of resistivity is enhanced due to the decrease of grain size.

Resistivity of both the samples has also been measured following a slightly different protocol namely FC_W as mentioned earlier. It has been found from the Fig. 3 (c) and (d) that the measured resistivity at zero field is strongly dependent on the magnetic field at which the samples were cooled before the measurement. There is a huge difference observable for NGSMO_N sample between the FC_W_8T curve and the ZFC_W curve (Fig. 3 (c)). Even at low field, namely, the FC_W_1T curve differs significantly with the ZFC_W curve. The difference increases with the increasing field. The NGSMO_B sample shows similar behaviour as depicted in the Fig. 3 (d). Only difference is that the bifurcation between the FC_W and ZFC_W curves is smaller compared to that of the NGSMO_N sample. These measurements elaborates the fact that the electronic phase of the sample is frozen or arrested during field cooling (a low resistive state) and does not come back to the original state even when the magnetic field is removed. Importantly, the reduction of grain size enhances this phenomenon.

At this point, the obvious question arises whether the frozen or arrested phase evolves with time and tries to come back to the original state as soon as the field is switched off. To have a clear picture about this, we have performed the time dependence of resistivity of both the bulk and nanometric samples. The samples were first cooled from a high temperature greater than the ferromagnetic (FM) - paramagnetic (PM) (as well as metal-insulator) transition temperatures under a highest magnetic field of 8 T down to the desired measuring temperature. After reaching the desired temperature the field was switched off and the resistivity of the samples were measured as a function of time for a sufficiently long period. This experiment has been performed at several temperatures for the NGSMO_N sample. Quite interestingly, it is found from the inset of Fig. 4 that the resistivity of the NGSMO_N and NGSMO_B samples increases with time. Resistivity shows a sharp rise initially and a slow variation thereafter for NGSMO_N sample at all temperatures. At an intermediate temperature (30 K) the change of resistivity is observed to be maximum indicating a non-monotonic dependence of the relaxation rate with temperature. Comparing the data at 3 K we find that increment of resistivity is higher for the NGSMO_N sample compared to NGSMO_B sample within the measured time span. These measurements confirm that the arrested electronic phase actually have a dynamic character and evolves with time. Thereafter, we try to find out a possible functional form of this time dependence of resistivity. Interestingly, we have observed that the relaxation is of logarithmic type similar to relaxation of magnetization found in many magnetic systems (spin glasses and magnetic nanoparticle systems) as evident from the linear nature of the curves in Fig. 4. The curves are fitted assuming an empirical relation $\rho(t) = \rho(0) + C \log(t)$ [C is a constant], successfully except the initial fluctuating part as shown in Fig. 4. Only at high temperature (60 K) we see some deviation from this law.

The magnetoresistance (MR) of the samples have been measured as a function of temperature as well as field. From Fig. 5 (a) it is evident that the magnetoresistance reaches nearly 100% in the low temperature regime at the highest applied field (8T) for both the samples and decreases as the temperature is increased. At 1 T field there is a marked difference between the MR vs. T curves of the samples at the low temperature regime. The NGSMO_N sample shows a higher MR which can be considered as an improvement of MR in the low temperature regime at low applied fields as an outcome of reduction of grain size. One can also notice a mutual crossover of the MR vs. T curves of the samples at both the applied fields below certain temperature. At the low temperature regime the nanometric sample shows higher MR, whereas, above a temperature (nearly the metal-insulator transition temperature) the bulk sample shows higher MR. The field dependence of MR and the effect of field cycling at different temperatures of the nanometric and bulk samples are shown in the Fig. 5 (b) to (f). The main observed feature of these measurements is that the resistive state after application of field does not return to the zero field state even when the field is swept to the negative highest value through zero. This is nothing but the similar arrest of phase on application of field as mentioned earlier. Noticeably, on increase of temperature from 3 K to 70 K in case of NGSMO_N sample this phenomenon is gradually weakened as the final value of resistance comes more close to the initial value. At 70 K we do not see any irreversibility at al. It is also found that the phenomenon is stronger in the nanometric sample when we compare the results with the bulk counterpart at 3 K (Fig. 5 (b) and Fig. 5 (f)). In order to have an insight about the mechanism of observed magnetoresistance we have tried to fit our experimental virgin MR vs. H data with different available models. Firstly, the spin dependent hopping model [20] has been employed to both the samples. According to this model the MR in the ferromagnetic regime is proportional to $B$(x), whereas MR is proportional to $B^2$(x) in the paramagnetic regime. $B$(x) is the Brillouin function. By tentatively assuming a mixture of ferromagnetic and paramagnetic phases in the sample the total MR can be expressed as MR = F(T)$B$(x) + P(T)$B^2$(x),where, F(T) and P(T) are the ferromagnetic and paramagnetic phase fraction in the sample, $x = g\mu_B J(T)H/k_B T$. The symbols have their usual significance. The MR vs. H curves of NGSMO_B sample is found to fit satisfactorily with this model. From the obtained values of J(T), the average value of total spin of a magnetic cluster, we have calculated the size of the cluster in terms of perovskite lattice unit (l.u.) as a function of temperature and is shown in the Fig. 6 (a). This model is found to provide a very poor fitting in case of NGSMO_N sample and not discussed here. Even if the fitting seems good for NGSMO_B sample one can observe the very large error bars in the values of calculated cluster size. This indicates that the above mentioned model is not sufficient to account for the magneto-transport mechanism in this sample entirely. Furthermore, we have applied a second model namely spin polarized tunneling model proposed by Raychaudhuri et al. [21]. According to this phenomenological description of MR, low field MR is mainly due to the spin polarized tunneling between adjacent grains and known as spin polarized tunneling MR ($MR_{SPT}$). MR at higher field comes from the double exchange mechanism and known as the intrinsic MR ($MR_{INT}$). The total MR at field H is given by , $MR = -\int_0^H [A\exp(-Bk^2) + Ck^2\exp(-Dk^2)]dk$ ('$MR_{SPT}$') $-JH-KH^3$ ('$MR_{INT}$') where, A, B, C, D, J and K are adjustable fitting parameters. This equation too, seems to fit well to the experimental data of NGSMO_B sample. From the obtained values of the parameters we have calculated the spin polarized tunneling MR and the intrinsic MR as a function of temperature and plotted in the Fig. 6 (b). The spin polarized tunneling MR is found to be the maximum near the metal-insulator transition and decreases as the temperature is increased. Intrinsic MR follows an opposite nature to the former. In case of NGSMO_N sample we again see a poor fitting. This clearly indicates that the magneto-transport mechanism is much more complicated for NGSMO_N sample compared to its bulk counterpart.

## 3. Magnetic properties

In order to probe the magnetic properties of the samples dc magnetization has been measured with temperature at very low dc applied fields following zero field cooled (ZFC)-field cooled (FC) protocols. The dc magnetic susceptibility ($\chi_{dc}$ = M/H) is plotted as a function of temperature in the Fig. 7 (a) for both the samples. The NGSMO_B sample shows a sharp transition at around 100 K, similar to a FM-PM transition. The FM-PM transition temperature as well as the metal-insulator transition temperature in NGSMO_B sample is much lower than that in the parent compound $Nd_{0.7}Sr_{0.3}MnO_3$ [1]. However, below the transition temperature a huge bifurcation between the ZFC-FC susceptibility is observed. This indicates that the ferromagnetic phase is not like a usual long range ordered ferromagnetic one, rather there could be a sufficient amount of frustration and disorder originating from antiferromagnetic coupling between Mn and Gd moments in this ferromagnetic state [8,11]. It is also noticeable that the ZFC curve of NGSMO_B shows a sharp increase in the value with increase of temperature in the lowest temperature regime. This is possibly originating from the ordering of the $Nd^{3+}$ moments [22]. The NGSMO_N sample shows an overall feature similar to that of its bulk counterpart. However, the PM-FM transition is found to broaden and flattened, whereas, the low temperature rise of moment is not found to be prominent. The destabilization of the ferromagnetic phase around the disorder surface region of the nano grains and the distribution of grain size can be attributed to the broadening of the magnetic transitions in NGSMO_N. The magnetization of both the samples has also been measured as a function of magnetic field at different temperatures and shown in the Fig. 7 (b). All the samples show a slightly non-saturating nature of the magnetization even up to 7 T field indicating an incomplete and progressive alignment of the magnetic moments of magnetic species in the samples. We have theoretically calculated the effective magnetic moment of the sample by adding the spin only moments ($g\sqrt{[S(S+1)]}$ $\mu_B$ ($g = 2$) of the magnetic ions. Firstly, assuming that the magnetism is only coming from B-site (Mn ions), the effective moment of our sample should be 4.6 $\mu_B$ / formula unit (f.u.). From the M-H data of the NGSMO_B sample at the lowest temperature (5 K) it is quite clear that the A-site magnetic ions ($Nd^{3+}$ and $Gd^{3+}$) contribute to the sample magnetization, the saturation magnetization ($M_s$ = 6.65 $\mu_B$ /f.u., calculated from approach to saturation model, $M = M_s - a/H + b/H^2$) being higher than this theoretical value. This observed moment (6.65 $\mu_B$ /f.u.) is also higher than the experimentally observed moment of $Nd_{0.7}Sr_{0.3}MnO_3$ [22]. This clearly indicates that the dopant $Gd^{3+}$ moment contributes to the magnetism of the sample and all the magnetic moments are in a partially aligned state. This observation is consistent with some earlier reports which showed that small amount of doping with magnetic rare earth material in the A- site actually enhances the net saturation moment of the sample [9]. The Curie - Weiss plot ($1/\chi$ vs. T) shows a straight line in the paramagnetic regime for NGSMO_B sample (Inset of Fig. 7 (a)). From the fitting of the equation $\chi = C/(T-\theta_P)$ (C is Curie -Weiss constant and $\theta_P$ is paramagnetic Curie temperature) to the experimental data we have calculated the value of C = 9.0 emu mole$^{-1}$Oe$^{-1}$ K and $\theta_P$ = 111 K. This experimental value of C corresponds to an effective value of the paramagnetic moment ($p_{eff}$) of 8.46 $\mu_B$ for the sample. Now assuming that all the magnetic ions in the sample ($Nd^{3+}$, $Gd^{3+}$, $Mn^{3+}$ and $Mn^{4+}$) are free in the paramagnetic regime, the theoretically calculated effective paramagnetic moment of this sample is $\sqrt{(\Sigma z p_{eff}^2)}$ = 6.74 $\mu_B$ (z = fractional occupancy of the magnetic ions) [11,23]. This value is lower than the experimental value. Such higher value of experimental $p_{eff}$ can be explained in terms of presence of FM clusters in the paramagnetic regime of the sample. This indicates that the FM-PM phase transition is not absolutely sharp and the presence of FM clusters, which have a higher spin moment than those of the individual ions, in the PM regime, gives rise to a higher value of effective paramagnetic moment [23]. Furthermore, we have tested the FM ordering in the light of Rhodes - Wohlfarth criteria [24]. In the localized model, $q_c/q_s$ = 1, where $p_{eff} = \sqrt{[(q_c (q_c+2)]}$. Here ½ $q_c$ is the effective spin number and $q_s$ is the average spontaneous moment (in $\mu_B$) calculated from M-H measurement. In the itinerant model, $q_s$ may be less than $q_c$ and $q_c/q_s$ (>1) gives a measure of the degree of itinerancy. We see, in our case, that $q_c$ = 7.52 $\mu_B$ and is greater than $q_s$ (= 6.65 $\mu_B$). This conforms to the fact that this system is an itinerant ferromagnet (due to double exchange mechanism). The NGSMO_N sample, on the other hand, always show smaller value of magnetization at all measured temperatures compared to the bulk sample (Fig. 7 (b)). It is expected as the magnetically disordered surface of the nano grains will contribute less to the total magnetization compared to the core region. The most striking difference observable from the M-H data is the change in coercivity with the change of grain size. At the lowest temperature (5 K), the coercivity of the NGSMO_B sample is found to be 160 Oe, whereas, it is as high as 1100 Oe for the NGSMO_N sample at the same temperature as shown in the inset of Fig. 7 (b). This effect can be attributed to the extraordinarily enhanced surface anisotropy due to the presence of defects, broken bonds, non stoichiometry etc. which produces number of pinning centers for the spins around the surface regime in the nanometric sample [25].

In order to obtain a clearer picture about the magnetic state of the system, we have performed the ac magnetic susceptibility (linear and nonlinear) measurements on both the samples as shown in the Fig. 8 (a) and (b). The frequency dependent real part of linear susceptibility ($\chi_1^R$) of the bulk sample is shown in the Fig. 8 (a). The $\chi_1^R$ vs. T curves for NGSMO_B sample shows similar features like the ZFC magnetization vs. T behavior with a FM-PM like transition at around 100 K. However, it is evident that below the transition the susceptibility shows a sharp drop unlike a conventional FM. More interestingly, the peaks of the $\chi_1^R$ vs. T curves are clearly frequency dependent as evident from the top right inset of Fig. 8 (a). The magnitude of the susceptibility gradually decreases and the position of the peak shifts to the high temperature on increase of frequency. These collectively signify a spin glass like behavior. To further investigate this feature we have measured the second and third harmonics (nonlinear susceptibility) of ac susceptibility as a function of temperature. The real part of second harmonic susceptibility ($\chi_2^R$) of NGSMO_B sample measured at different frequencies is shown in the Fig. 9 (a). The $\chi_2^R$ vs. T curves show a sharp peak at around the FM-PM transition. The second harmonic (and other even harmonics) is generally observed when symmetry breaking spontaneous magnetization is present in the sample [26]. The observed peak in $\chi_2^R$ signifies the presence of ferromagnetism in this sample. Interestingly, the peak in $\chi_2^R$ also shows a frequency dependence like the linear susceptibility (the peak shifts to the higher temperature with the increase of frequency as shown in the inset of Fig. 9). This can be regarded as a very unconventional feature as in a canonical spin glass $\chi_2$ is not present but $\chi_1$ shows frequency dependence at the freezing temperature of the sample. In our NGSMO_B sample, on the other hand, a spin glass like frequency dependent peak in $\chi_1$ has been observed, whereas, $\chi_2$ is also found to be present and exhibits a peak at the same transition temperature with similar frequency dependence. Keeping in mind these behaviours in the ac susceptibility (ferromagnetism along with glassiness) of NGSMO_B sample, we label the FM state of this sample as a glassy ferromagnet (GFM). A strong test for a spin glass like behaviour in a system is the measurement of third order susceptibility ($\chi_3$) which shows a sharp negative peak at around the freezing temperature and diverges in the low field limit ($h_{ac} \rightarrow 0$) [27]. In this regard, a thorough measurement of $\chi_3$ and its analysis has been carried out on our samples. The temperature dependent real part of third order susceptibility ($\chi_3^R$) of NGSMO_B sample is shown in the Fig. 10 (a) at different ac magnetic fields ($h_{ac}$). In the low field regime $\chi_3^R$ shows a sharp peak at around the GFM-PM transition temperature supporting the glassy behavior of the system. Now we plot the peak/maximum values of $\chi_3^R$ after normalizing it with respect to that at $h_{ac}$ = 15 Oe, as a function of magnetic field ($h_{ac}$) as shown in Fig. 10 (b). Interestingly, $\chi_3^R$ for NGSMO_B sample shows a diverging nature as the ac field is reduced showing a spin glass like feature. One noticeable fact is that increase of $h_{ac}$ changes the nature of variation of $\chi_3^R$ vs. T considerably. As the ac field is increased $\chi_3^R$ shows a crossover nature (clearly visible for $h_{ac}$= 5 Oe), precisely, a positive peak below the transition temperature and a negative peak above it, unlike a single negative peak in the low $h_{ac}$ regime. This cross over nature is a characteristic feature of a long range ordered magnetic system and generally observed near FM-PM transition temperature of a conventional FM [28]. Possibly, when the magnetic field is high enough all the spins in a grain are forced to respond to the field coherently and displays a long range ordered behavior instead of exhibiting a spin glass like criticality. So, on the basis of features observed from the measurement of $\chi_3^R$ it is clear that the system (NGSMO_B) has a ferromagnetic character accompanied with glassiness. Finally, we try to analyze the frequency dependence of $\chi_1^R$ employing known models. Application of Arrhenius law of the form $\tau = \tau_0 \exp (E_a/\kappa_B T_f)$ [29,30] yields unphysical values of $E_a$ and $\tau_0$ showing inapplicability of this law to the observed phenomena. It also hints at the fact that the system is not a non interacting particle system. However, the relaxation process could not be explained by Vogel-Fulcher law also. Finally, application of critical slowing down model of the form $\tau = \tau_0 (T_f / T_g-1)^{-z\upsilon}$ [29,30] produces physically meaningful set of parameters. Here, $\tau = 1/f$, $T_g$ = spin glass transition temperature, $z\upsilon$ = critical exponent, $\tau_0$ = the characteristic time and $T_f$ = frequency dependent peak of $\chi_1^R$ vs. T curves. On application of this model (as shown in the inset of Fig. 8(a)) we get $T_g$ = 86 K, $z\upsilon$ = 3 and $\tau_0 = 10^{-17}$ s. The value of $z\upsilon$ falls on the lower side of the observed values (4-12) for the spin glass materials [29]. However, the value of $\tau_0$ is smaller than the reported values of this parameter ($10^{-11}$s - $10$-$^{14}$s) in spin glasses [29]. This signifies that in spite of having spin glass like signatures, our system is different from a canonical spin glass.

In a similar manner, we have also measured the linear and non-linear ac susceptibility of the nanometric NGSMO_N samples. The linear susceptibility ($\chi_1^R$) of the sample at different frequencies as a function of temperature is plotted in the Fig. 8 (b). There is clearly no shift of the peaks with frequency is observable. The FM-PM transition is found to be very broad compared to the bulk sample. The variation of $\chi_2^R$ in the inset of Fig. 9 (b) and $\chi_3^R$ in the inset of Fig.10 (a) with temperature show peaks around the FM-PM transition. All the higher order susceptibilities are found to be frequency independent. A comparative study of the divergence of $\chi_3^R$ (Fig. 10 (b)) around the transition temperature with the bulk one shows that unlike NGSMO_B sample, $\chi_3^R$ trends to saturate as the magnetic field $h_{ac}$ is reduced towards zero for nanometric NGSMO_N sample. From the above mentioned study on ac susceptibilities of the NGSMO_N sample it is quite clear that the glassy phase is destabilized in it due to the reduction of grain size.

## IV. Discussion

Now we turn to describe the possible scenario behind the observed features of the system. It has been mentioned earlier that introducing $Gd^{3+}$ replacing $Nd^{3+}$ decrease the average A- site ionic radii, increase the A -site disorder and induces extra magnetic coupling with the Mn site. Decrease of <$r_A$> can directly be related to the decrease of FM-PM transition temperature of $Nd_{0.4}Gd_{0.3}Sr_{0.3}MnO_3$ compared to the parent compound $Nd_{0.7}Sr_{0.3}MnO_3$. The above mentioned effects on doping of Gd collectively enhance the phase separation of the system. Though it is almost impossible to get an idea of the actual microscopic nature of the phase separated state, one can easily put forward a possible picture of the magnetic and electronic state in the sample. Keeping in mind the observed metal insulator transition and ferromagnetic as well as antiferromagnetic interactions in the system we can divide the system in broadly two classes of phase. One of them is the low resistive metallic phase which magnetically should have ferromagnetic character allowing double exchange to operate. The other one is high resistive insulating phase and possibly have a less ferromagnetic or antiferromagnetic character and even be a canted spin state. Previously, the possibility of a separate spin glass phase coexisting with the ferromagnetic state was reported [10]. Even though our system shows a spin glass like behaviour, we believe that a separate SG phase is not possible because the system (NGSMO_B) does not show any other spin glass transition except the FM-PM transition. Rather the magnetic state of the system is itself glassy in nature as a whole. However, as the temperature of the system is increased PM clusters (insulating in nature) develops, grow in size and the sample undergoes the FM-PM transition. Similarly, the metal insulator transition can be explained. In the low temperature regime metallic phase dominates and as the temperature is increased insulating phases grow in size leading to an insulating nature of the system above a certain temperature (metal-insulator transition temperature). In such an assumption of a two component system the resistivity/conductivity can be conveniently described through effective medium approximation [31,32]. The effective resistivity ( $\rho_{eff}$ ) of the system is given by,

$$f(T,H)\left[\frac{\rho_{FM} - \rho_{eff}}{2\rho_{FM} + \rho_{eff}}\right] + \left[1 - f(T,H)\right]\left[\frac{\rho_{AFI} - \rho_{eff}}{2\rho_{AFI} + \rho_{eff}}\right] = 0$$

Here $f$ represents the volume fraction of the ferromagnetic metallic phase whose resistivity is $\rho_{FM}$ and (1- $f$ ) is the volume fraction of antiferromanetic insulating phase whose resistivity is $\rho_{AFI}$. Obviously, $\rho_{FM}$ is less than $\rho_{AFI}$. In a simpler approach where these two phases are connected in series the effective resistivity can be written as,

$$\rho_{eff} = f(T,H)\rho_{FM} + \left[1 - f(T,H)\right]\rho_{AFI}$$

So increase of $f$ (or decrease of (1- $f$ )) will decrease the resistivity of the system and vice versa. Necessarily, we take the phase fraction $f$ as function of temperature and field. This naturally explains the observed properties of our system.

It is a well known issue that large MR of manganites cannot be explained on the basis of double exchange alone. One has to incorporate the phase separation scenario [33]. It is quite obvious that the effect of phase separation has an important role to play in our system showing high negative MR even in low magnetic field. On application of magnetic field the volume fraction of the low resistive metallic phase ( $f$ ) is increased (and (1- $f$ ) is decreased) and gives rise to the negative magnetoresistance. Moreover, when the field is switched off the latest phase separated state tends to persists (with only a slow evolution with time). This is reflected in the huge bifurcation between the ZFC-W and FC-W resistivity curves. It seems that the electronic phase gets arrested when the sample is field cooled. A similar bifurcation in resistivity of manganite showing first order phase transition has been reported [18,34]. We note that our nanometric NGSMO_N sample, in which the maximum history dependence is found, exhibits a prominent first order phase transition. Recently, similar phenomena have been observed in some other manganite materials and the effect is attributed to a more commonly used terminology called kinetic arrest [18, 34-36]. When we look at the ZFC-FW and FC-FW curves, the maximum bifurcation is found at an intermediate field. When the field is high both the ZFC-FW and FC-FW state tend to saturate to the high field low resistive state (saturation of $f$ ) reducing the bifurcation with the increase of magnetic field.

The observed time dependence of FC resistivity can be explained taking the thermal energy into account. On switching off the field, the insulating clusters starts growing in size increasing the overall fraction of the insulating state in the sample and a simultaneous increase in the resistivity with time. At the lowest temperature, obviously the stability of the individual phases is very high through their individual barrier energies. As the temperature is increased, thermal energy comes to play, competes with the energy barriers and enhances the time rate of change of resistivity. At much higher temperature, however, thermal energy dominates and disturbs the individual character of the phases and reflected in a less rate of change of resistivity. The reduction of the irreversibility in the MR vs. H curves on increase of temperature (Fig. 5) can be attributed to the thermal instability of the phases in a similar fashion. All these observations of thermo-magnetic irreversibility of resistivity in the system indicate that $f(T,H)$ and [1- $f(T,H)$ ] are not only function of temperature and field but also irreversible with respect to these parameters.

To explain the magnetic behaviour of the system, one has to incorporate the role of $Gd^{3+}$ necessarily. We know that the bulk parent compound $Nd_{0.7}Sr_{0.3}MnO_3$ does not show any kind of glassy behaviour. However, presence of $Gd^{3+}$ in the sample creates the extra antiferromagnetic exchange interaction between Gd and Mn ions. We also know that the exchange between $Mn^{3+}$ and $Mn^{4+}$ is ferromagnetic in nature. So, presence of two types of exchange bonds as well as random distribution of this bonds provide two necessary ingredients, frustration and disorder, for a sample to display spin glass like behaviour.

However, our system (NGSMO_B) is different from a canonical glass and exhibits comparably a faster dynamics evidenced from the smaller value of $\tau_0$.

On reduction of grain size the phase separation scenario discussed will be modified. The nanograins can be visualized as crystallographically ordered core with a shell or surface where defects, broken bonds etc. may exists in large numbers. Moreover, surface to volume ratio increases when the size is reduced. We believe that on such a background (nanosize grains) the spatial extent of the individual phases will be reduced which in turn enhance the phase separation in the sample. Effectively, we observe the enhanced magnetoresistance, time dependence of resistivity and thermo-magnetic irreversibility in the NGSMO_N sample. However, due to the destabilization of the phases, ultimately the glassy phase ceases to exist in NGSMO_N. The lack of required number of interacting spins over a certain region in NGSMO_N destabilizes the glassy behaviour which is necessarily a collective behaviour of a spin system via multiple exchange interaction.

## V. Conclusions

In summary, the effect of doping of $Gd^{3+}$ in $Nd_{0.7}Sr_{0.3}MnO_3$ has been investigated in terms of measurement of electronic and magnetic properties. We have observed that the system show thermo-magnetic history dependence in its electronic- and magneto-transport properties. Both the bulk and nanometric samples show high magnetoresistance even in the low field. The temporal dependence of resistivity is investigated after a field cooling and resistivity is found to increase with time when the field is switched off. Magnetically, the bulk system shows a spin glass like behaviour along with a ferromagnetic nature. We have designated this magnetic state as a glassy ferromagnetic state. The observed phenomena have been explained through the phase separation scenario in manganites. The reduction of grain size is found to enhance the history dependence in transport properties but destabilizes the glassy behaviour. This work clearly represents the strong influence of doping as well as size reduction on manganite material.

**Acknowledgement**

One of the authors (T. K. Nath) would like to acknowledge the financial assistance of Department of Science and Technology (DST), New Delhi, India through project no. IR/S2/PU-04/2006.

## Figure captions

**Fig. 1.** (a) Plot of experimental high resolution x-ray diffraction data of the samples (dots) with the fitted curve (line) and the difference on Rietveld refinement. (b) High resolution transmission electron microscopy image of the NGSMO_N sample. (c) Distribution of grain size of NGSMO_N sample with the fitted curve with lognormal distribution function. (d) Field emission scanning electron microscopy image of the NGSMO_B sample. (e) Selected area diffraction pattern of the NGSMO_N sample.

**Fig. 2.** Resistivity (ρ) vs. temperature (T) plot of the (a) NGSMO_N and (b) NGSMO_B sample at H=0 T. Inset of (a) shows the enlarged version of the curve for NGSMO_N sample. Inset of (b) shows the fitted low temperature resistivity data of NGSMO_B sample.

**Fig. 3.** The variation of ZFC_FW and FC-FW resistivities at different magnetic fields of (a) NGSMO_N and (b) NGSMO_B samples with temperature. The variation of ZFC_W and FC_W resistivities at different fields of (c) NGSMO_N and (d) NGSMO_B samples with temperature.

**Fig. 4.** The variation of normalized resistivities with logarithmic time scale along with the fitted curves (lines) assuming a relation $\rho(t) = \rho(0) + C \log(t)$. The inset shows the variation of normalized resistivities of the samples with time at different temperatures after field cooling at 8 T field.

**Fig. 5**. (a) Plot of magnetoresistance (MR) of the samples as a function of temperature at H = 1 T and H= 8 T. Panels (b) to (e) show the MR of NGSMO_N sample as a function of magnetic field (-8T- 0T - 8T) at different temperatures. Panel (f) displays the variation of MR with magnetic field of NGSMO_B sample at 3 K.

**Fig. 6.** (a) The plot of ferromagnetic and paramagnetic cluster size of NGSMO_B sample as obtained from spin dependent hopping model as a function of temperature. (b) The plot of intrinsic ($MR_{INT}$) and spin polarized tunnelling ($MR_{SPT}$) magnetoresistance as a function of temperature of NGSMO_B sample as obtained from spin polarized tunnelling analysis.

**Fig. 7.** (a) The variation of the dc magnetization of the samples under zero field cooled (ZFC) and field cooled (FC) protocols. Inset of (a) shows the inverse of dc susceptibility as a function of temperature of NGSMO_B sample. The line represents the fitted curve on application of Curie - Weiss law.(b) Magnetization vs. magnetic field plot of the samples at different temperatures showing hysteresis behaviour of them. Inset of (b) shows the magnified image of the M-H curves to clarify the variation of coercivity in nanometric and bulk systems.

**Fig. 8.** (a) The variation of the real part of linear ac susceptibility ($\chi_1^R$) of NGSMO_B sample with respect to temperature at different frequencies and at ac applied field ($h_{ac}$) of 3 Oe. Top

right inset shows the magnified version of the main plot around the transition region. Top left inset shows the fitted curve to the experimental data under the scheme of critical slowing down model. (b) Plot of $\chi_1^R$ vs. temperature of the NGSMO_N sample at different frequencies at $h_{ac}$ = 3 Oe.

**Fig. 9.** (a) Plot of real part of the second harmonic of ac susceptibility ($\chi_2^R$) of NGSMO_B sample as a function of temperature at different frequencies. Inset shows the magnified image of the peak showing the frequency dependence. (b) The $\chi_2^R$ vs. temperature plot of the NGSMO_N sample.

**Fig. 10.** (a) Variation of the real part of third harmonic of ac susceptibility ($\chi_3^R$) of NGSMO_B sample as a function of temperature at different ac fields. Inset of (a) shows the variation of the same for NGSMO_N sample with temperature at a particular field (2 Oe) and frequency (555.3 Hz). (b) Plot of normalized peak value of $\chi_3^R$ (with respect to peak value of $\chi_3^R$ at $h_{ac}$ = 15 Oe) of both the samples as a function of ac field.

## References

1. Y. Tokura, Colossal Magnetoresistive Oxides, Gordon and Breach Science, Singapore,2000.
2. C. Zener, Phys. Rev. **82,** 440 (1951).
3. H.Y. Hwang et al., Phys. Rev. Lett.**75**, 914 (1995).
4. K.R. Mavani et al, Appl. Phys. Lett. **86**, 162504 (2005).
5. N. Ghosh et al., Phys. Rev. B **70**, 184436 (2004).
6. R. Mathieu et al., Phys. Rev. B **74**, 020404(R) (2006).
7. K. F. Wang et al., Appl. Phys. Lett. **89**, 222505 (2006).

8. Y. Sun et al., Phys Rev. B **66**, 094414 (2002).
9. L. E. Hueso et al., J. Magn. Magn. Mater **238**, 293 (2002).; T. Terai et al. Phys. Rev. B **58**, 14908 (1998).
10. S. C. Bhargava et al., Phys. Rev. B **71**, 104419 (2005).
11. J. Hemberger et al., Phys. Rev. B **70**, 024414 (2004).
12. P. Dey et al., J. Phys.: Condens. Matter. **19**, 376204 (2007).
13. M. K. Srivastava et al., J. Appl. Phys. **107**, 09D726 (2010).
14. R. K. Pati et al., J. Am. Ceram. Soc. **84**, 2849 (2001).
15. Ll. Balcells et al., Phys. Rev. B **58**, R14697 (1998).
16. P. K. Muduli et al., J. Appl. Phys. **105**, 113910 (2009).
17. S. Mukhopadhyay et al., J. Phys.: Condens. Matter **21**, 186004(2009).
18. R. Rawat. et al., J. Phys.: Condens. Matter **19**, 256211 (2007).
19. M. A. Manekar et al., Phys. Rev. B **64**, 104416 (2001).
20. P. Wagner et al., Phys. Rev. Lett. **81**, 3980 (1998).
21. P. Raychaudhuri et al., J. Appl. Phys. **84**, 2048 (1998).
22. J. Park et al., J. Kor. Phy. Soc. **36**, 412 (2000).
23. R. P. Borges et al., Phys. Rev. B **60**, 12847 (1999).
24. P. Rhodes et al., Proc. Roy. Soc. London, Ser. A 273, 247 (1963); E. P. Wohlfrath et al., J. Magn. Magn. Mater. **7**,113 (1978).
25. E. Winkler et al., Phys. Rev. B **72**, 132409 (2005).
26. A. K. Pramanik et al., J. Phys. Condens. Matter **20**, 275207 (2008).
27. A. Bajpai et al., Phys. Rev. B **55**, 12439 (1997); M. Suzuki et al., Prog. Theor. Phys **58**, 1151 (1977).
28. S. Nair et al., Phys. Rev. B **68**, 094408 (2003).
29. J. A. Mydosh, Spin Glasses: An Experimental Introduction, CRC Press (1993).

30. K. Binder et al., Rev. Mod. Phy. **58**, 801 (1986).

31. S. Krikpatrick, Phys. Rev. Lett. **27**, 1722 (1971).

32. S. Ju et al., J. Phys.: Condens. Matter **14**, L631(2002).

33. E. Dagotto, Nanoscale Phase Separation and Colossal Magnetoresistance, Springer (2003).

34. A. A. Wagh et al., J. Phys. Condens. Matter **22**, 026005 (2010).

35. P. Chaddah et al., Phys. Rev. B **77**, 100402(R), (2008).

36. K. Kumar et al., Phys. Rev. B **73**, 184435 (2006).

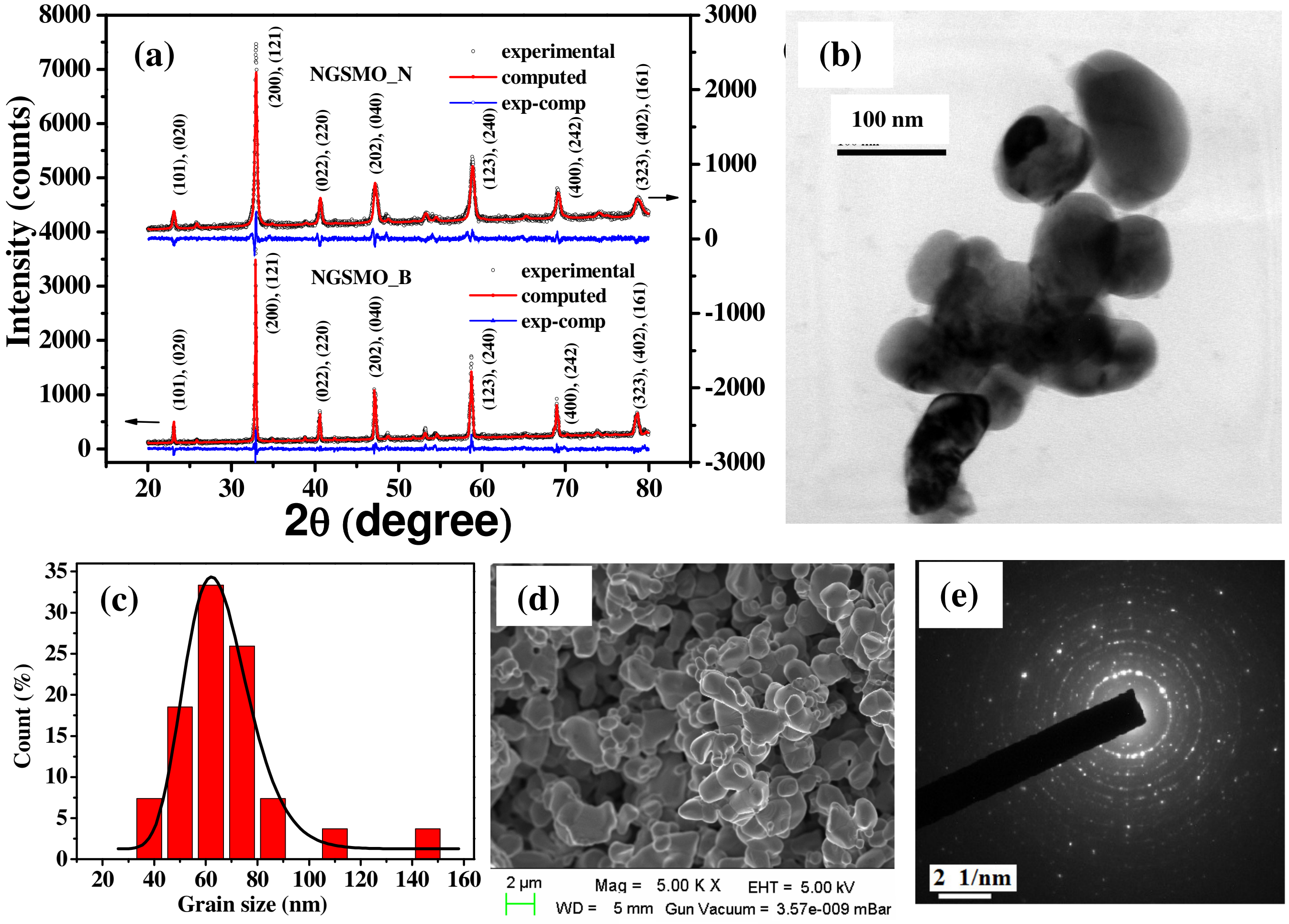

# Fig. 1, S. Kundu et al.

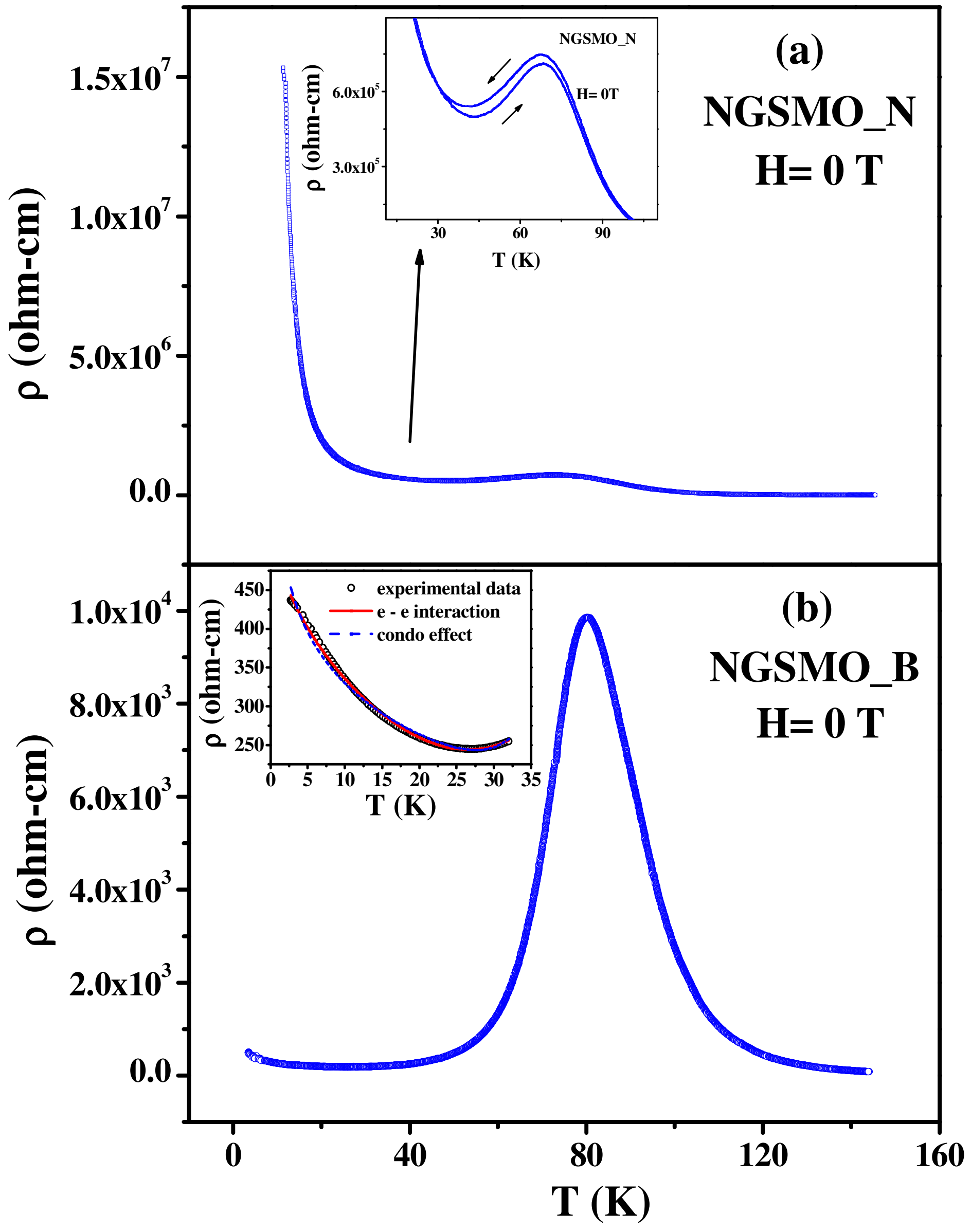

# Fig. 2, S. Kundu et al.

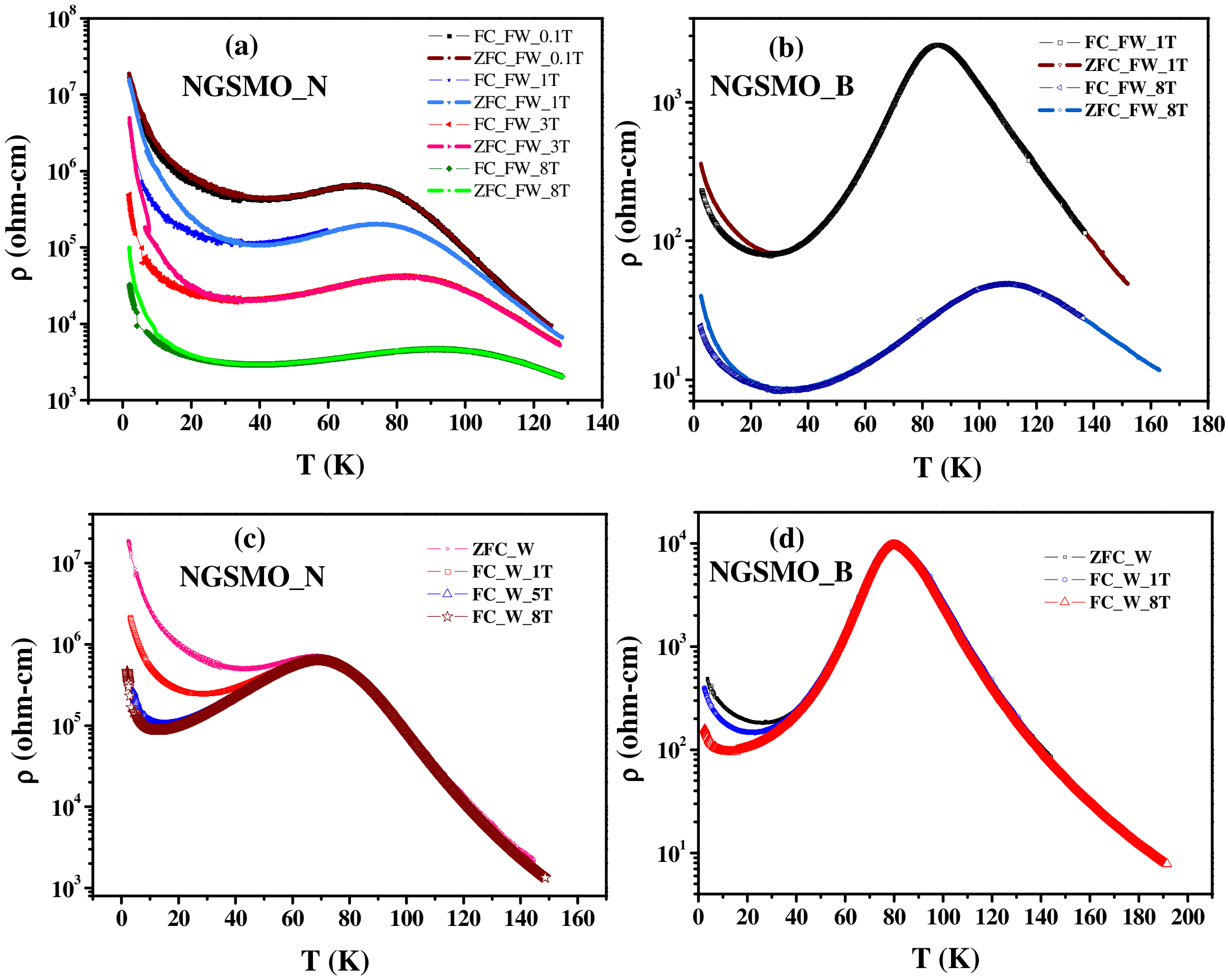

# Fig. 3, S. Kundu et al.

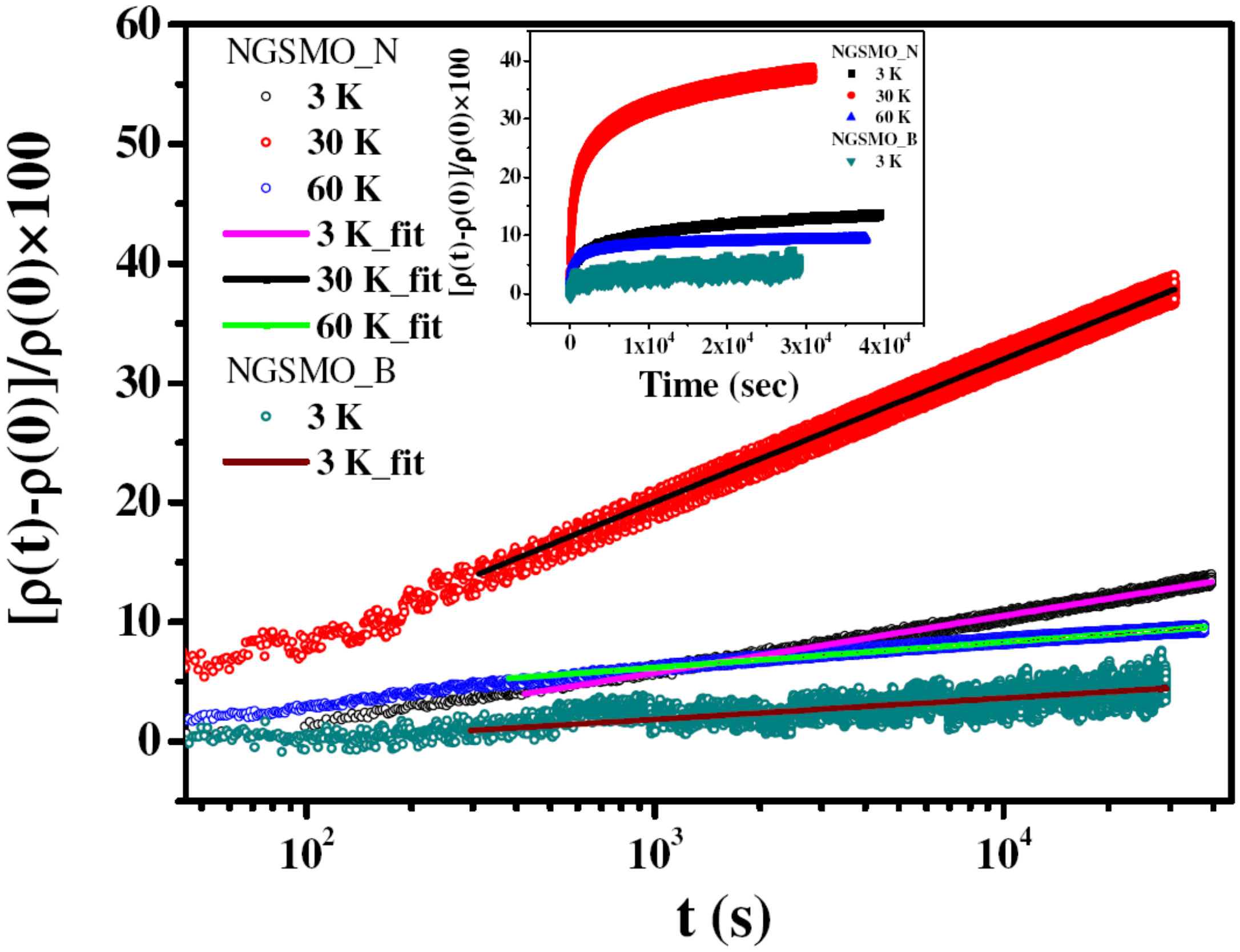

**Fig. 4, S. Kundu et al.**

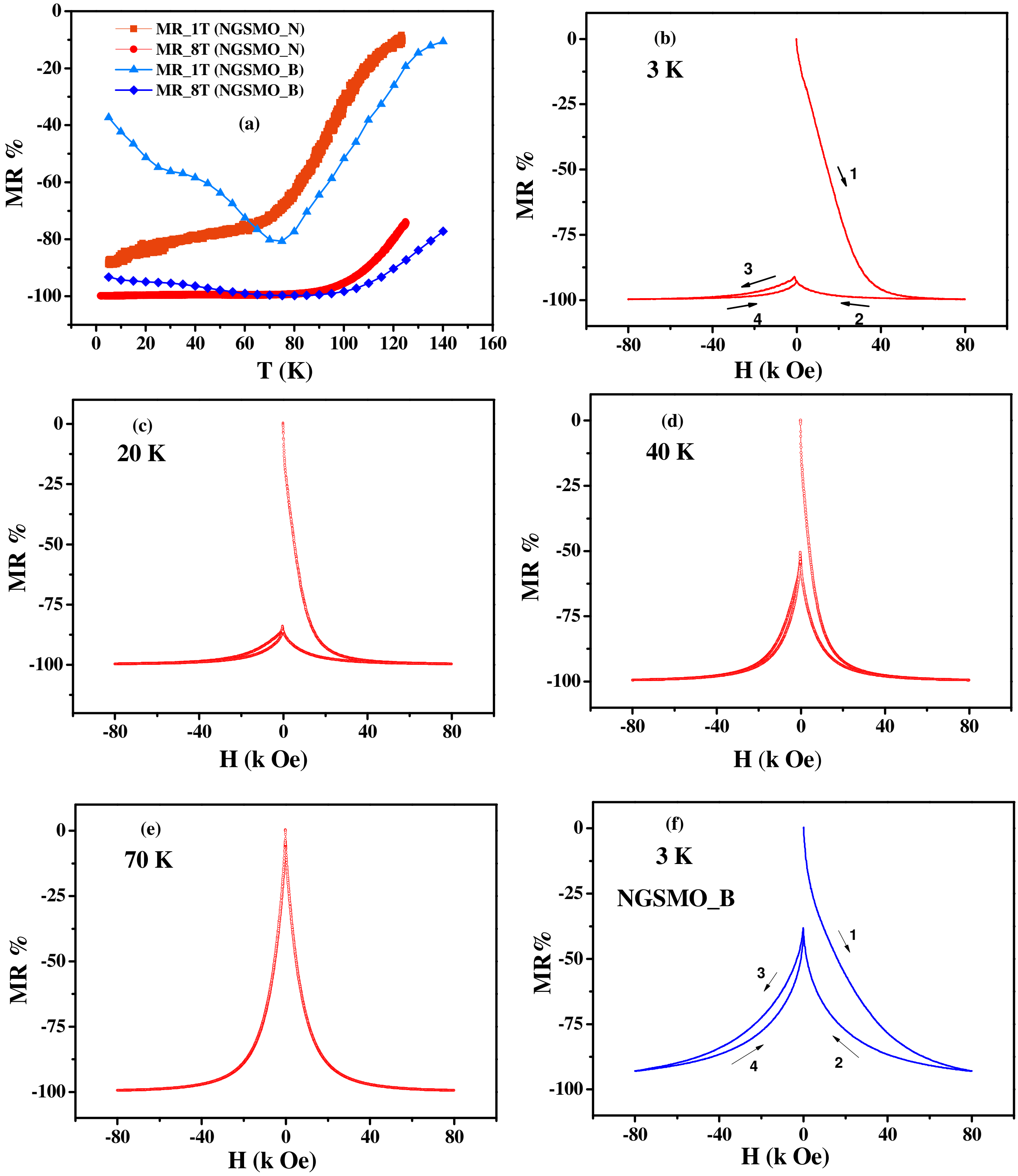

# Fig. 5, S. Kundu et al.

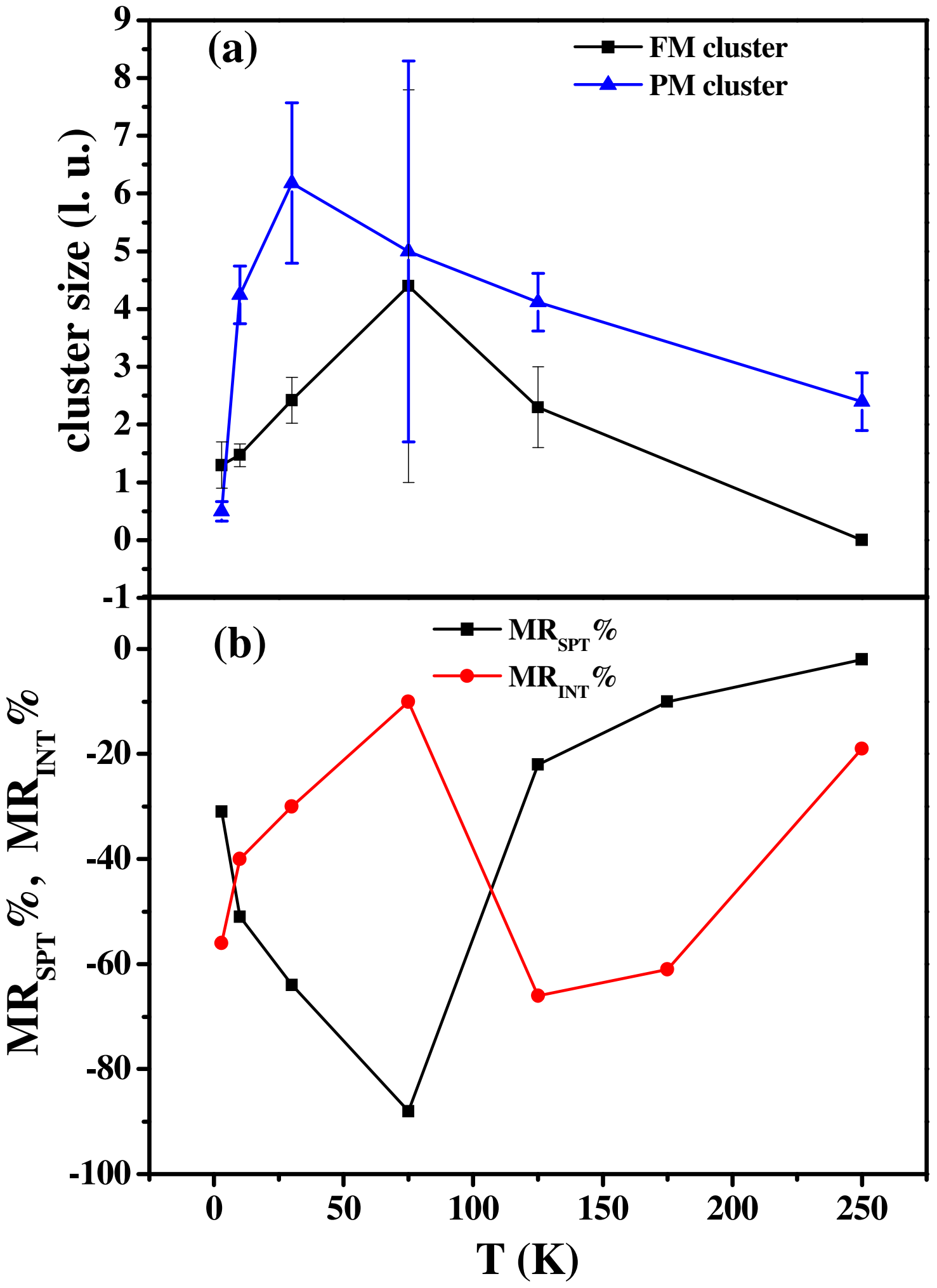

# Fig. 6, S. Kundu et al.

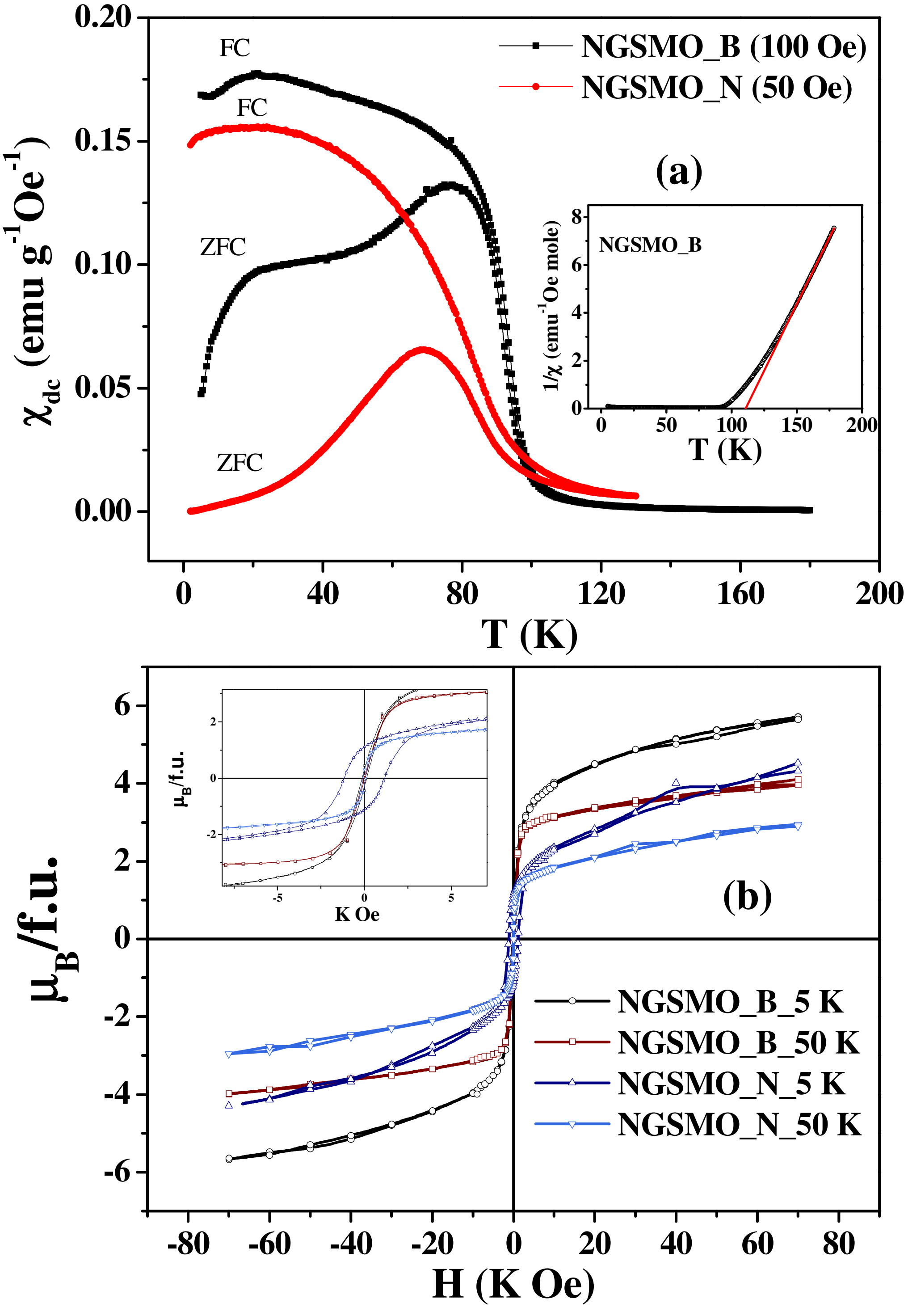

**Fig. 7, S. Kundu et al.**

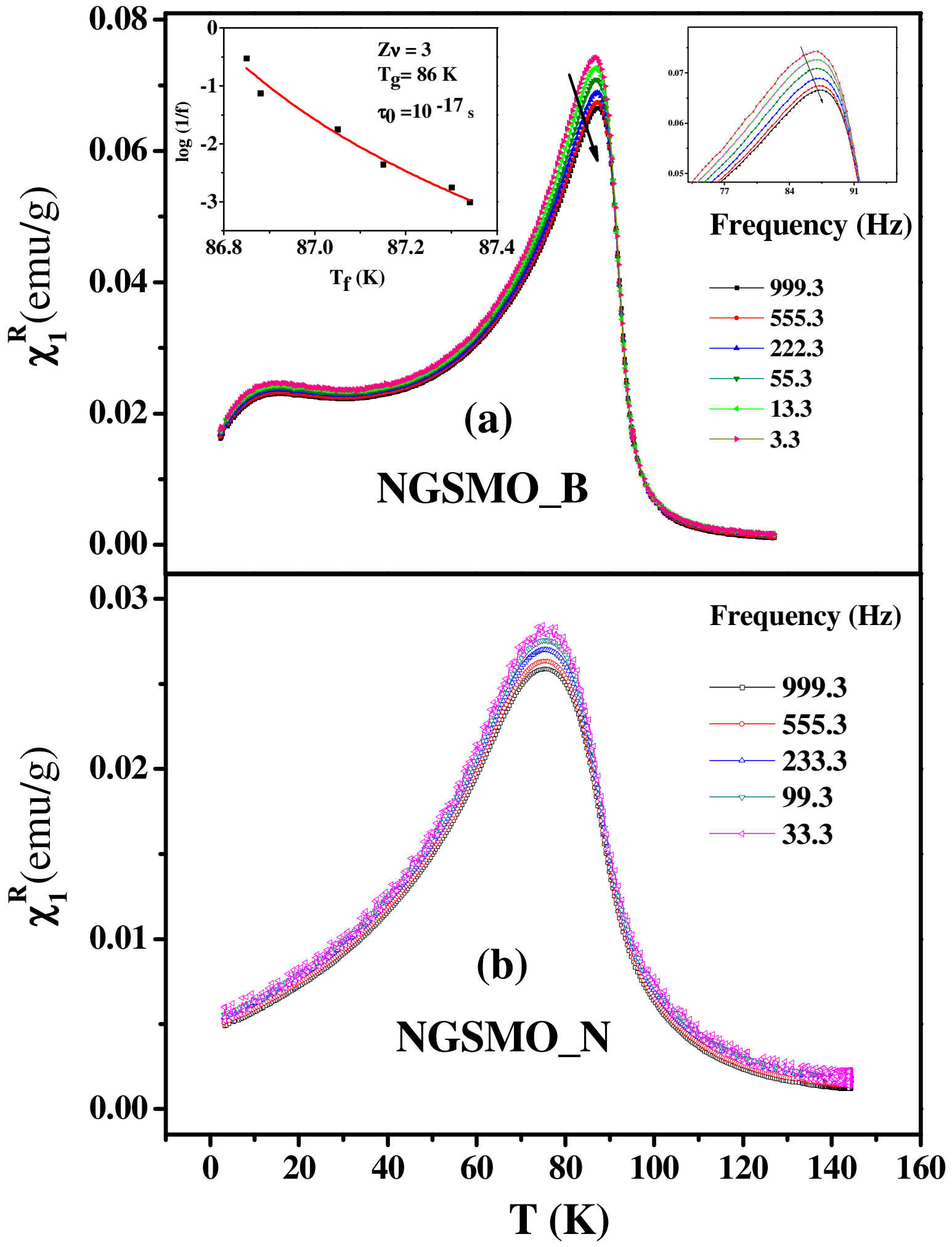

# Fig. 8, S. Kundu et al.

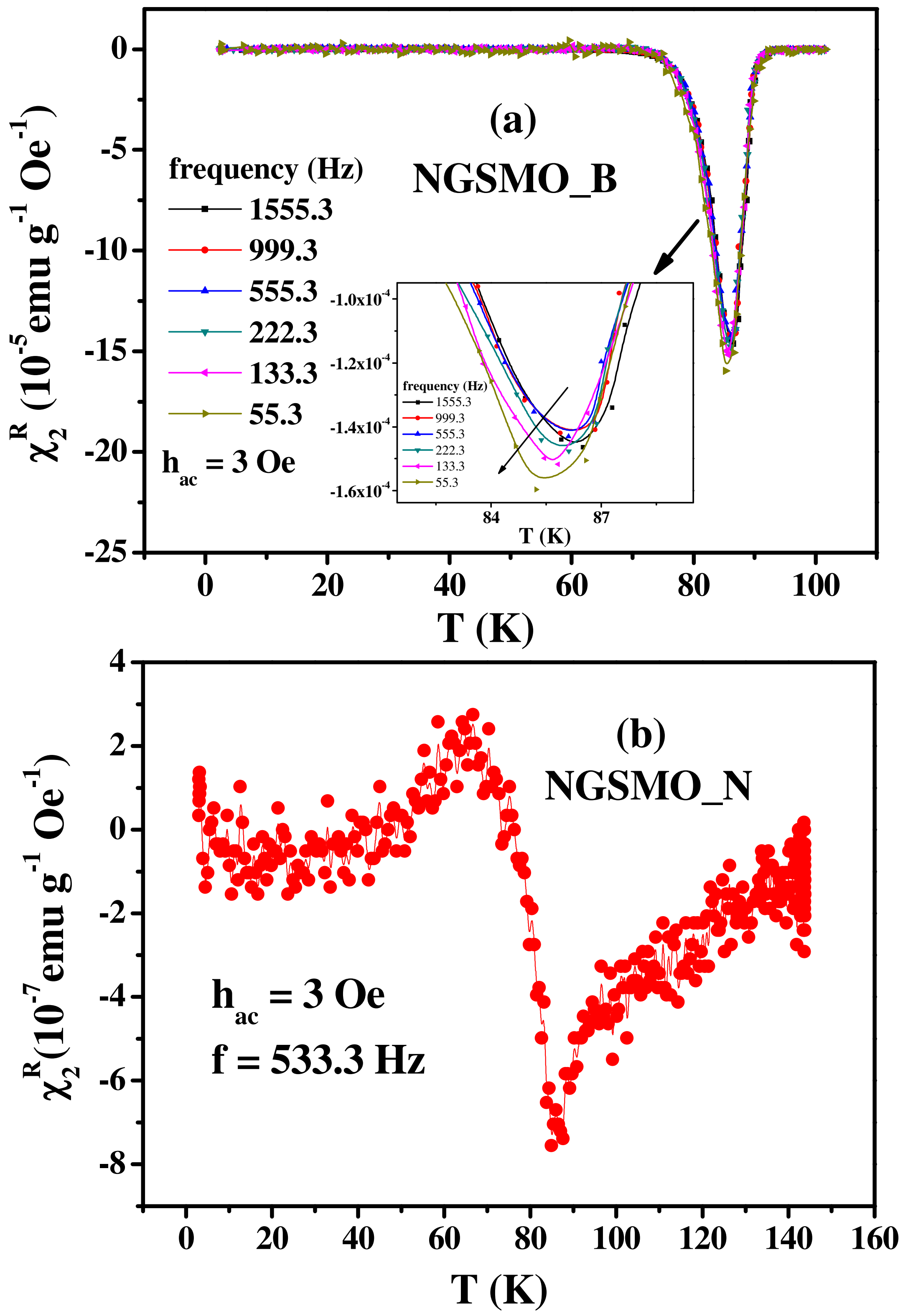

**Fig. 9, S. Kundu et al.**

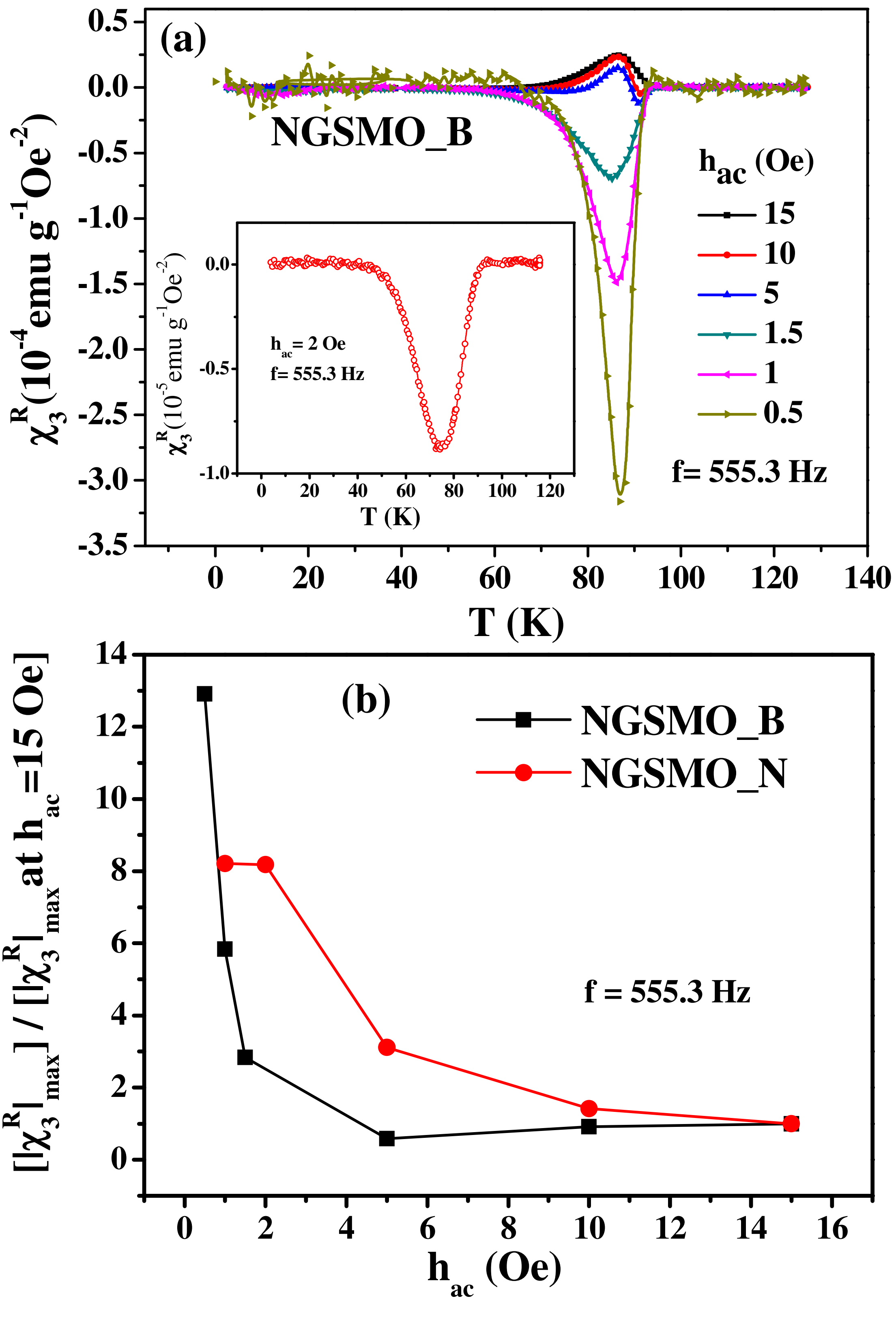

**Fig. 10, S. Kundu et al.**